\documentclass[floatfix,twocolumn,showpacs,preprintnumbers,amsmath,amssymb,pra,superscriptaddress,longbibliography]{revtex4-1}
\usepackage{color}
\usepackage[usenames,dvipsnames,svgnames,table]{xcolor}
\usepackage[colorlinks=true,linkcolor=blue,urlcolor=blue,citecolor=blue]{hyperref}
\usepackage{mathtools}
\usepackage{graphicx}
\usepackage{dcolumn}
\usepackage{array}
\usepackage{lipsum}
\usepackage{bm}
\usepackage{subfigure}
\usepackage{amssymb}
\usepackage{multirow}
\usepackage{tabularx}
\usepackage{amsmath}
\usepackage{braket}
\graphicspath{{plots/}}
 \usepackage{lipsum}
\usepackage{mathrsfs}

\begin{document}
\title{
Path Integral Monte Carlo Simulations of liquid $^3$He without Fixed Nodes:\\ Structural Properties and Collective Excitations
}

\author{Tobias Dornheim}
\email{t.dornheim@hzdr.de}

\affiliation{Center for Advanced Systems Understanding (CASUS), D-02826 G\"orlitz, Germany}

\affiliation{Helmholtz-Zentrum Dresden-Rossendorf (HZDR), D-01328 Dresden, Germany}

\author{Zhandos A.~Moldabekov}

\affiliation{Center for Advanced Systems Understanding (CASUS), D-02826 G\"orlitz, Germany}

\affiliation{Helmholtz-Zentrum Dresden-Rossendorf (HZDR), D-01328 Dresden, Germany}

\author{Jan Vorberger}
\affiliation{Helmholtz-Zentrum Dresden-Rossendorf (HZDR), D-01328 Dresden, Germany}

\author{Burkhard Militzer}
\affiliation{Department of Earth and Planetary Science, University of California, Berkeley, California 94720, USA}
\affiliation{Department of Astronomy, University of California, Berkeley, California 94720, USA}

\begin{abstract}
We present extensive new \emph{ab initio} path integral Monte Carlo (PIMC) simulations of normal liquid $^3$He without any nodal constraints. This allows us to study the effects of temperature on different structural properties like the static structure factor $S(\mathbf{q})$, the momentum distribution $n(\mathbf{q})$, and the static density response function $\chi(\mathbf{q})$, and to unambiguously quantify the impact of Fermi statistics. 
In addition, the dynamic structure factor $S(\mathbf{q},\omega)$ is rigorously reconstructed from imaginary-time PIMC data, and we find the familiar phonon-maxon-roton dispersion that is well known from $^4$He and has been reported previously for two-dimensional $^3$He films [Nature \textbf{483}, 576-579 (2012)]. The comparison of our new results for both $S(\mathbf{q})$ and $S(\mathbf{q},\omega)$ to neutron scattering measurements reveals an excellent agreement between theory and experiment.
\end{abstract}

\maketitle

Ultracold helium constitutes one of the most actively investigated quantum systems and has been of central relevance for our understanding of important physical concepts such as superfluidity~\cite{cep} and Bose-Einstein condensation~\cite{Yukalov2011}. Due to its nature as a strongly correlated quantum liquid, helium exhibits an intricate interplay of non-ideality effects, quantum statistics, and thermal excitations. Naturally, an accurate description of physical effects such as the lambda phase-transition in $^4$He must capture all of these effects simultaneously---a challenging task beyond simple mean-field models and perturbative approaches. 

This challenge was met by Feynman~\cite{feynman2010quantum} in terms of the path integral formalism, where the complicated quantum system of interest is exactly mapped onto an effective classical system of interacting ring-polymers~\cite{Chandler_Wolynes_JCP_1981}. Specifically, this quest for an accurate description of helium~\cite{Fosdick_original_PIMC} has given rise to the widespread path integral Monte Carlo (PIMC) method~\cite{Berne_JCP_1982,Takahashi_Imada_PIMC_1984,Pollock_Ceperley_PRB_1984}, one of the most successful tools in statistical physics, quantum chemistry, and related disciplines. While the ergodic sampling of the permutation-space, which is required to take into account the effect of quantum statistics, is rendered nontrivial by the strong repulsion between two He atoms at short range~\cite{Boninsegni2005}, this problem has been solved by the recent \emph{worm algorithm} idea~\cite{boninsegni1,boninsegni2}.

The PIMC method gives straightforward access to important physical observables like the superfluid fraction~\cite{ultracold2}, the momentum distribution~\cite{cep}, and the static structure factor, which has resulted in excellent agreement between theory and experiments for $^4$He; see the review by Ceperley~\cite{cep} for details. In addition, PIMC simulations can be used as the starting point for an analytic continuation~\cite{JARRELL1996133} giving access to the dynamic structure factor $S(\mathbf{q},\omega)$~\cite{Boninsegni_maximum_entropy,Boninsegni1996,Ferre_Boronat_PRB_2016}---a key quantity in neutron scattering experiments~\cite{PhysRev.113.1379,Skoeld_1980,PhysRevA.5.1377,Bramwell2014}. In particular, PIMC-based data for $S(\mathbf{q},\omega)$ have given important insight into the connection between superfluidity and quasi-particle excitations of a roton nature.

In stark contrast, the accurate PIMC simulation of $^3$He is substantially hampered by the notorious fermion sign problem~\cite{dornheim_sign_problem,troyer}, which leads to an exponential increase of the computation time with increasing the system size $N$ or decreasing the temperature $T$. Therefore, Ceperley has used PIMC within the uncontrolled \emph{fixed-node approximation}~\cite{Ceperley1991} to present the first results for $^3$He. Moreover, this investigation was restricted to the total energy, and the agreement to experimental data~\cite{Greywall_PRB_1983} was inconclusive. In the meantime, other PIMC investigations of $^3$He have been sparse~\cite{DuBois}, and, to our knowledge, no data have been presented for either the structural properties or the spectrum of collective excitations.

This is unfortunate, as ultracold $^3$He offers a potential wealth of interesting physical effects. First and foremost, we mention the superfluid phase transition due to the formation of Cooper pairs in the range of $T\lesssim2.5$mK~\cite{vollhardt2013superfluid}.  In addition, it has been recently shown~\cite{Panholzer_Nature_2012,Nava_PRB_2013} that two-dimensional $^3$He exhibits a rich phonon-maxon-roton dispersion relation that phenomenologically resembles the more well-known dispersion of $^4$He. At the same time, we note that the experimental investigation of bulk $^3$He is notoriously difficult~\cite{GUCKELSBERGER1989681}, and a thorough theoretical approach is, thus, indispensable to capture the underlying physical mechanisms.



In this Letter, we remedy this unsatisfactory situation by carrying out extensive direct PIMC simulations of normal liquid $^3$He  \emph{without any nodal restrictions}. Therefore, our simulations are exact, but computationally extremely costly when the temperature is decreased, cf.~the discussion of Fig.~\ref{fig:Sign} below. This allows us to present highly accurate results for the temperature dependence of important properties such as the static structure factor $S(\mathbf{q})$, the momentum distribution function $n(\mathbf{q})$, and the static density response function $\chi(\mathbf{q})$. Furthermore, we are able to unambiguously characterize the impact of Fermi statistics onto these properties, which is comparably small for $S(\mathbf{q})$ and $\chi(\mathbf{q})$, but very pronounced on $n(\mathbf{q})$ in the small-momentum range.

In addition, we compute the imaginary-time density--density correlation function $F(\mathbf{q},\tau)$ for the same parameters, which gives us access to the dynamic structure factor $S(\mathbf{q},\omega)$. First and foremost, we indeed find the familiar phonon-maxon-roton dispersion relation~\cite{Panholzer_Nature_2012} in these spectra, which is qualitatively similar to normal liquid $^4$He~\cite{Ferre_Boronat_PRB_2016} at similar conditions. In addition, our new PIMC data for the spectrum of collective excitations are in excellent agreement to measurements from neutron scattering experiments~\cite{Skoeld_1980} where they are available, thereby leading to an unprecedented agreement between experiment and theory.

To our knowledge, this Letter reports the first comprehensive PIMC study of an ultracold atomic bulk system of fermions at finite temperature without the nearly ubiquitous \emph{fixed-node approximation}~\cite{Ceperley1991}, thereby opening up new avenues for the investigation of other applications such as quantum-dipole systems~\cite{Filinov_PRA_2016,Dornheim_PRA_2020}, bilayer structures~\cite{Neumann1356,Filinov_2009}, or isotopic mixtures of helium~\cite{Boninsegni_PRL_1995,Boninsegni_PRL_1997,Boninsegni_JCP_2018}.

\textbf{Results.} All PIMC results that are shown in the present work have been obtained by using the extended ensemble approach that was introduced in Ref.~\cite{Dornheim_PRB_nk_2021}. In addition, we consider strictly spin-unpolarized $^3$He
at a number density $n=N/V=0.016355${\AA}$^{-3}$, and the convergence with the number of imaginary-time steps has been carefully checked; see the Supplemental Material~\cite{supplement} for additional details.

Let us start our investigation by touching upon the fermion sign problem, which constitutes the main computational bottleneck.
This is illustrated in Fig.~\ref{fig:Sign} where we show our PIMC results for the average sign $S$ (see e.g. Ref.~\cite{dornheim_sign_problem}) for $N=14$ and $N=38$ unpolarized $^3$He atoms interacting via the usual Aziz-2 potential~\cite{Aziz_JCP_1979}. In particular, $S$ constitutes a measure for the amount of cancellation of positive and negative terms in the simulations and monotonically decreases with $T$. Specifically, it holds $S=1$ in the high-temperature limit when the effect of quantum statistics vanishes, whereas $S\to0$ towards the ground state~\cite{krauth2006statistical}. Furthermore, it is well-known that the Monte Carlo error bar of an observable $\hat A$ scales as $\Delta A/A\sim 1/S$, resulting in a computational increase of $C=1/S^2$~\cite{dornheim_sign_problem}. This is shown in the bottom panel of Fig.~\ref{fig:Sign} and can be interpreted as follows: For $T=5$K, which is close to the Fermi temperature of $^3$He, the effect of quantum statistics is negligible and there is no increase in the computational effort, i.e., $C\sim1$. In contrast, we find $C\sim10^3$ for $T=2$K and $N=38$, which means that we need $1000$ times the compute time compared to a bosonic PIMC simulation without the sign problem. While this is still feasible on modern supercomputers with $\mathcal{O}(10^5)$ CPUh, this temperature constitutes the limit of the present investigation.

\begin{figure}\centering
\includegraphics[width=0.385\textwidth]{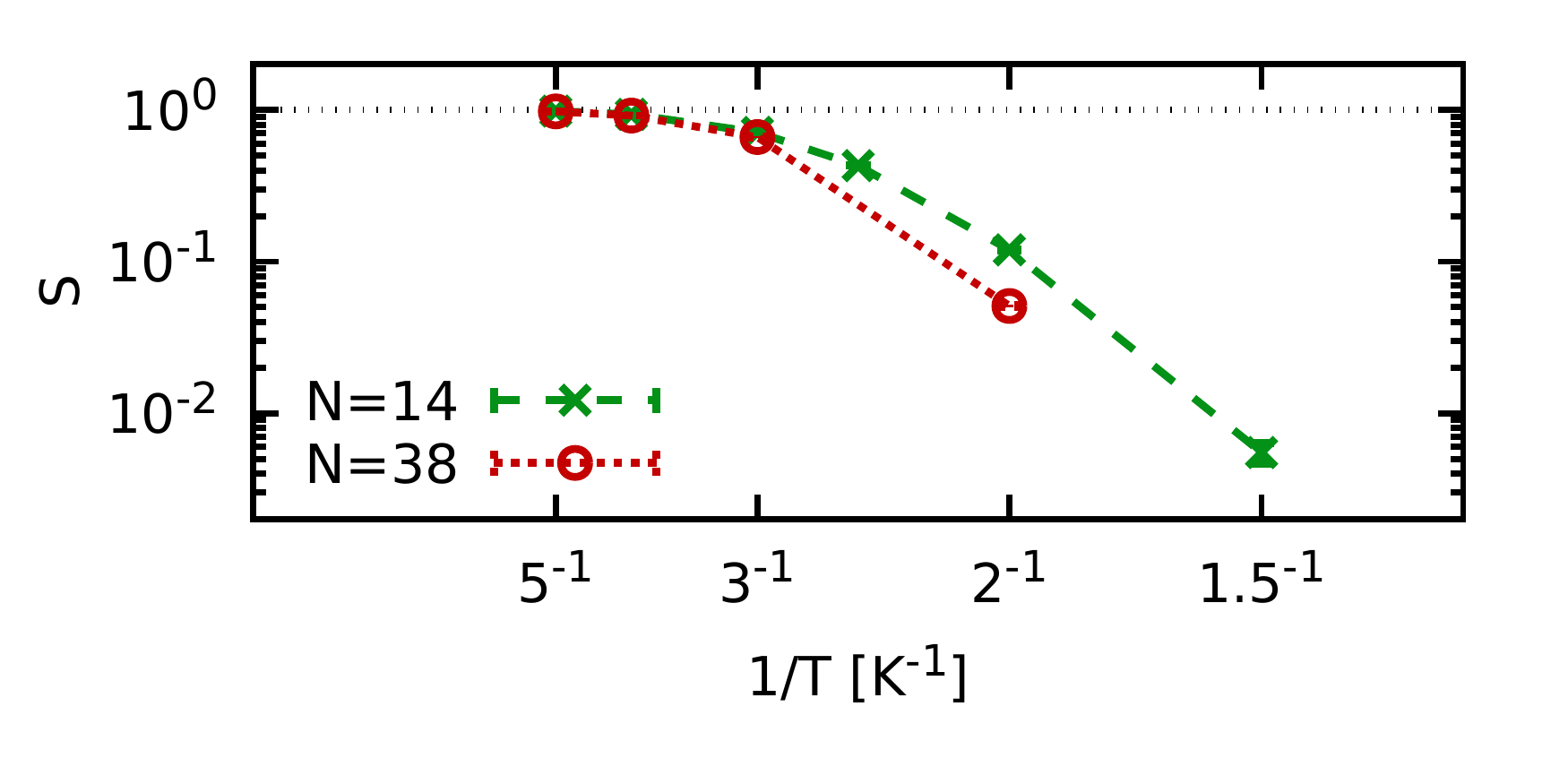}\\\vspace*{-1.15cm}
\includegraphics[width=0.385\textwidth]{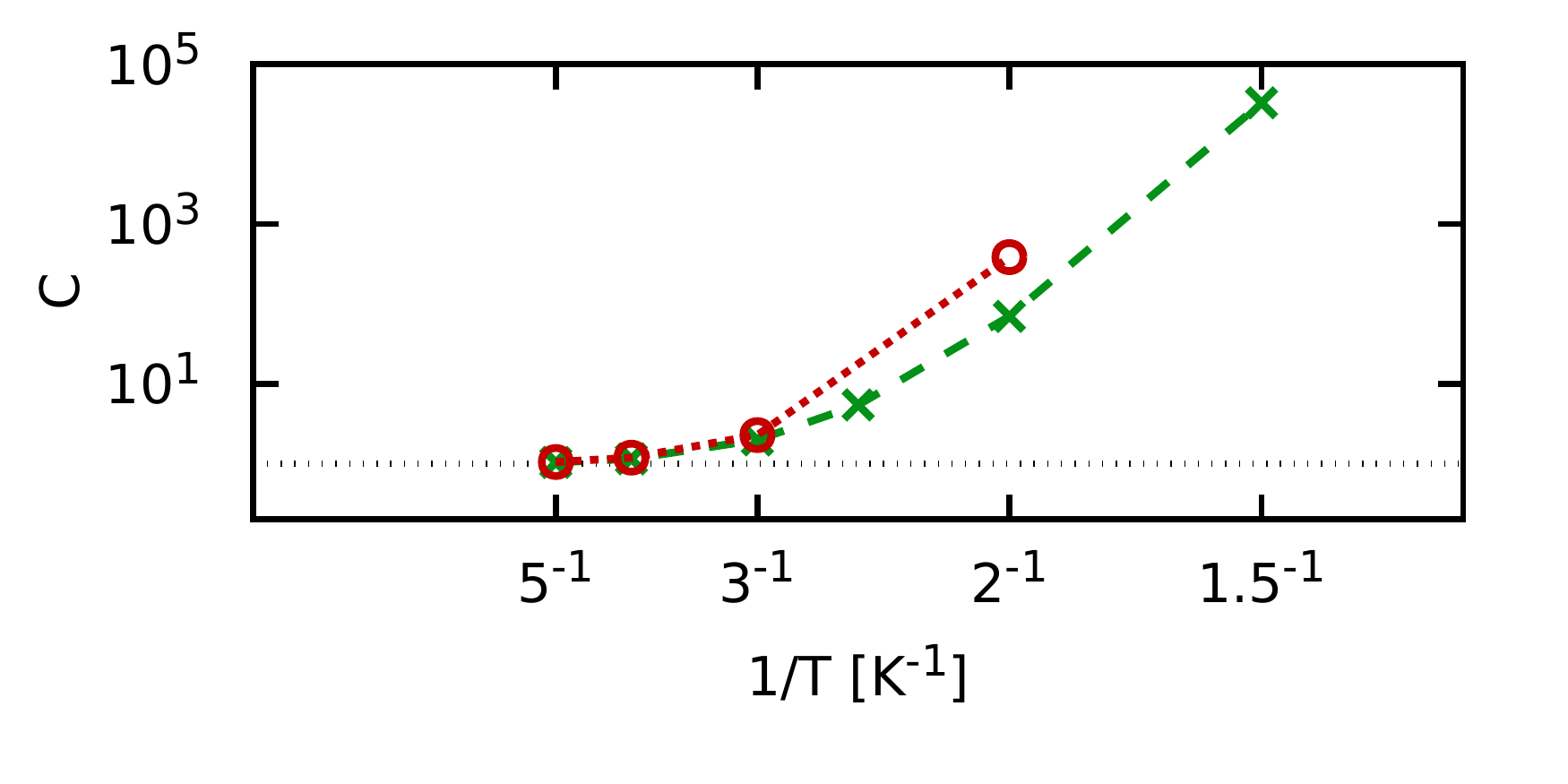}\\\vspace*{-0.5cm}
\caption{\label{fig:Sign}
Average sign $S$ (top) and computational increase $C=1/S^2$ (bottom) for $N=14$ and $N=38$ $^3$He atoms as a function of the inverse temperature $T^{-1}$.
}
\end{figure}

\begin{figure}\centering
\includegraphics[width=0.4485\textwidth]{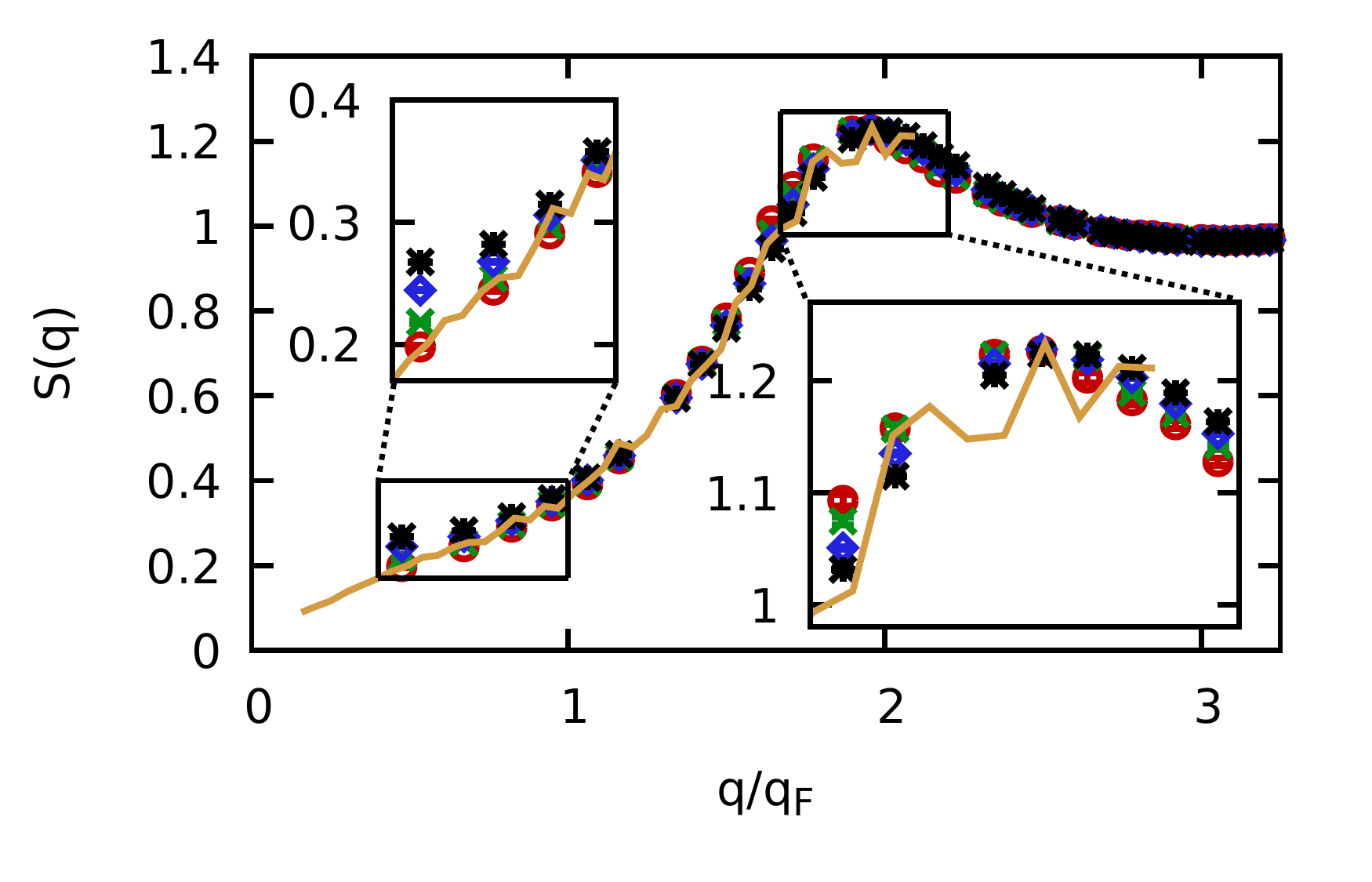}\\\vspace*{-1.3cm}
\includegraphics[width=0.4485\textwidth]{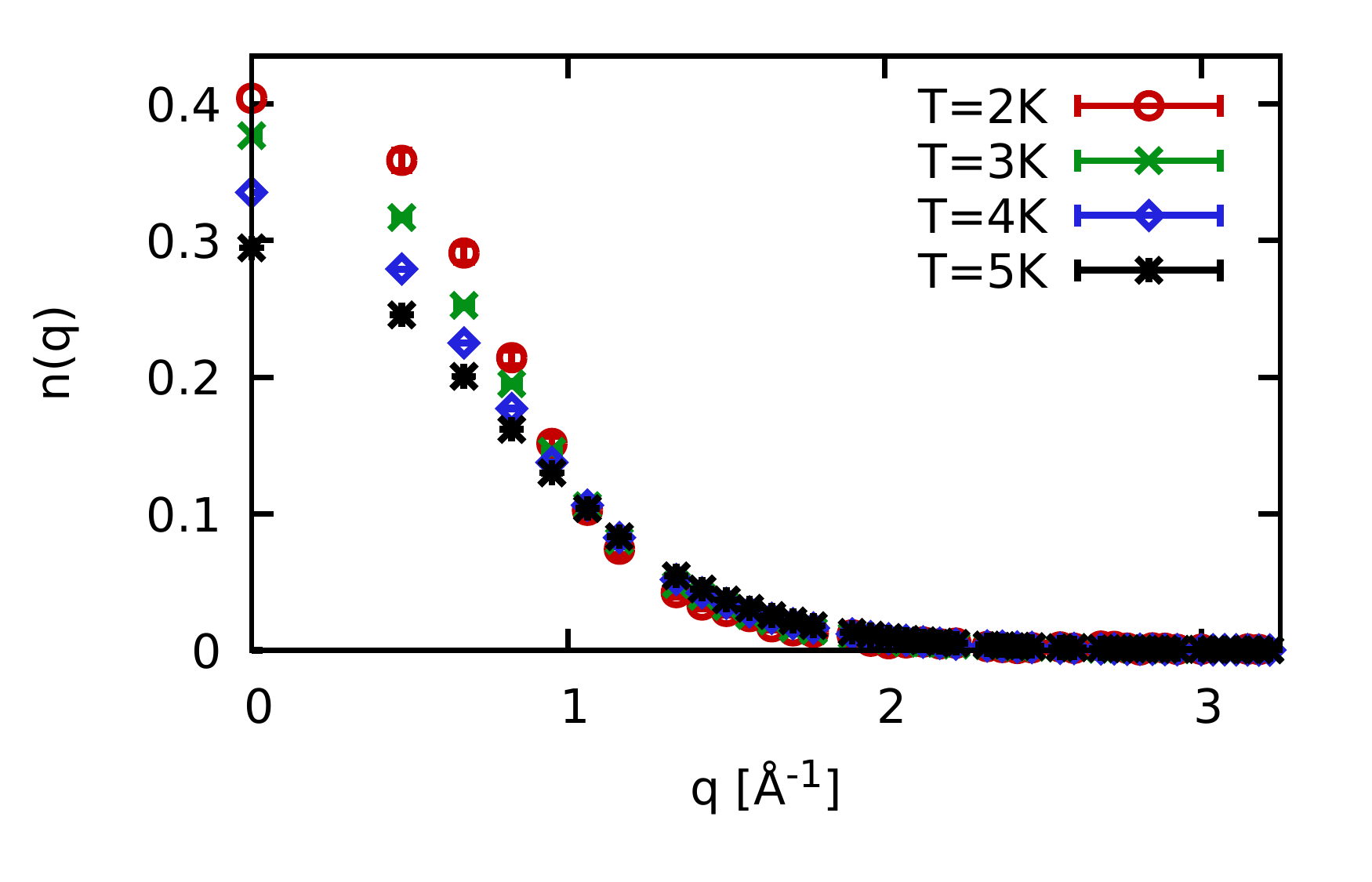}\\\vspace*{-1.3cm}
\hspace*{-0.132cm}\includegraphics[width=0.46\textwidth]{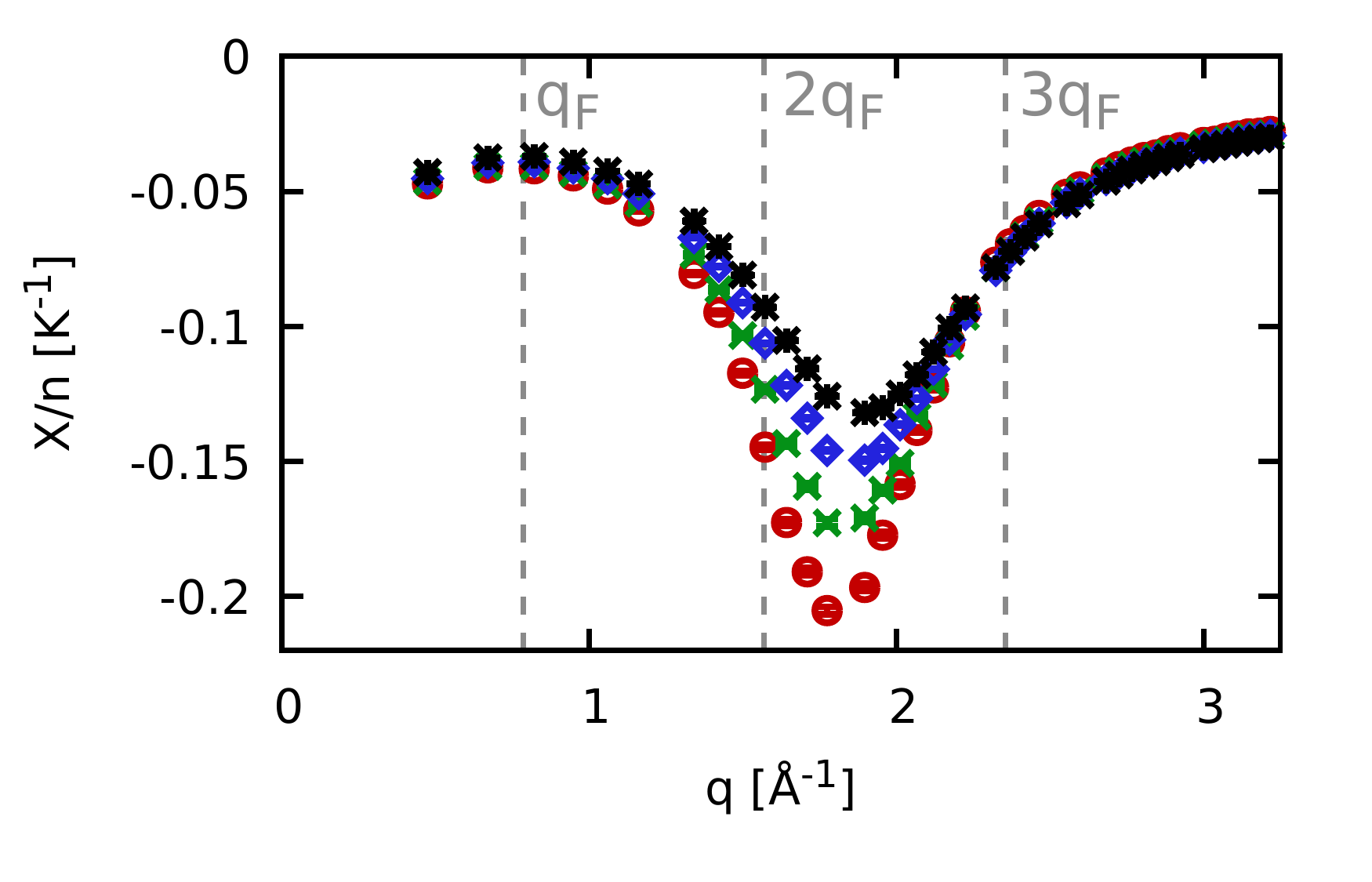}\\\vspace*{-0.5cm}

\caption{\label{fig:T_dependence}Static properties of $^3$He for different temperatures $T$:
Shown are PIMC results for the static structure factor $S(q)$ [top], momentum distribution function $n(q)$ [center], and static density response function $\chi(q)$ [bottom], see Eq.~(\ref{eq:static_chi}). Solid yellow: experimental data for $S(q)$ by Hallock~\cite{Hallock1972}.
}
\end{figure}

Let us next consider the temperature dependence of the structural properties of normal liquid $^3$He, which are depicted in Fig.~\ref{fig:T_dependence} for $T=5,4,3,2$K. The top panel shows our PIMC results for $S(\mathbf{q})$, which only exhibits a minor dependence on $T$. More specifically, the most pronounced temperature effect manifests in the long wavelength limit, which is determined by the isothermal compressibility~\cite{Ferre_Boronat_PRB_2016}, see also the inset showing a magnified segment. For completeness, we note that the exact $q\to0$ limit cannot be accessed in our simulations due to the finite simulation cell~\cite{dornheim_prl,dornheim_cpp}. Apart from this momentum quantization effect~\cite{dornheim_prl}, we find no finite-size effects in our PIMC results; see the Supplemental Material~\cite{supplement} for a corresponding analysis. Both the position and the shape of the peak are hardly affected by $T$, which is consistent to earlier findings for $^4$He~\cite{Ferre_Boronat_PRB_2016}.
The solid yellow line in the same panel shows experimental data for $S(\mathbf{q})$ at $T=0.41$K by Hallock~\cite{Hallock1972}, which is in excellent agreement to the PIMC data for the lowest temperature.

The center panel shows the same information for the momentum distribution function $n(\mathbf{q})$, which we have estimated following the procedure described in Ref.~\cite{Dornheim_PRB_nk_2021}. For this property, the temperature plays an important role as the $^3$He atoms are pushed towards larger momenta by thermal excitations. Furthermore, $n(\mathbf{q})$ does not resemble a step function even for the lowest depicted temperature, $T=2$K, corresponding to a reduced temperature of $\Theta=k_\textnormal{B}T/E_\textnormal{F}\approx0.4$. 

Lastly, the bottom panel corresponds to the static density response function~\cite{nolting}, which we estimate from the imaginary-time version of the fluctuation--dissipation theorem~\cite{Dornheim_JCP_ITCF_2021,bowen2},
\begin{eqnarray}\label{eq:static_chi}
\chi(\mathbf{q}) = -n\int_0^\beta \textnormal{d}\tau\ F(\mathbf{q},\tau) \quad ,
\end{eqnarray}
with the definition of the imaginary-time correlation function
\begin{eqnarray}\label{eq:F}
F(\mathbf{q},\tau) = \braket{\hat{n}(\mathbf{q},0)\hat{n}(-\mathbf{q},\tau)}\ ,
\end{eqnarray}
where $\hat{n}(\mathbf{q},\tau)$ is the density operator in Fourier space evaluated at $\tau\in[0,\beta]$; see also Ref.~\cite{Dornheim_JCP_ITCF_2021} for a generalization. We find that $\chi(\mathbf{q})$ exhibits an interesting, non-monotonous structure: i) both in the limits of large and small $q$, the response function only weakly depends on the temperature at these conditions. Further, $\chi(\mathbf{q})$ does not approach zero in the long wavelength limit, as there is no perfect screening~\cite{review,kugler_bounds} for helium due to the short-range nature of the effective two-body potential~\cite{Aziz_JCP_1979}; ii) the density response function exhibits a pronounced peak around $q\approx1.8${\AA}$^{-1}$, which corresponds to $q\approx2.25q_\textnormal{F}$ (where $q_\textnormal{F}$ is the Fermi wave number~\cite{quantum_theory}). In fact, this feature closely resembles recent fermionic PIMC results~\cite{dornheim_ML,dornheim_electron_liquid} for the density response of a uniform electron gas at warm dense matter conditions~\cite{new_POP} at similar values of the reduced temperature $\Theta$ and reduced wavenumber $x=q/q_\textnormal{F}$. Finally, the peak of the density response substantially depends on $T$. Specifically, the peak location is directly connected to the attractive minimum in the inter-atomic potential. Increasing $T$ leads to a more weakly correlated system, and, therefore, less collective behaviour, which manifests in a weaker density response.

\begin{figure}\centering
\includegraphics[width=0.4285\textwidth]{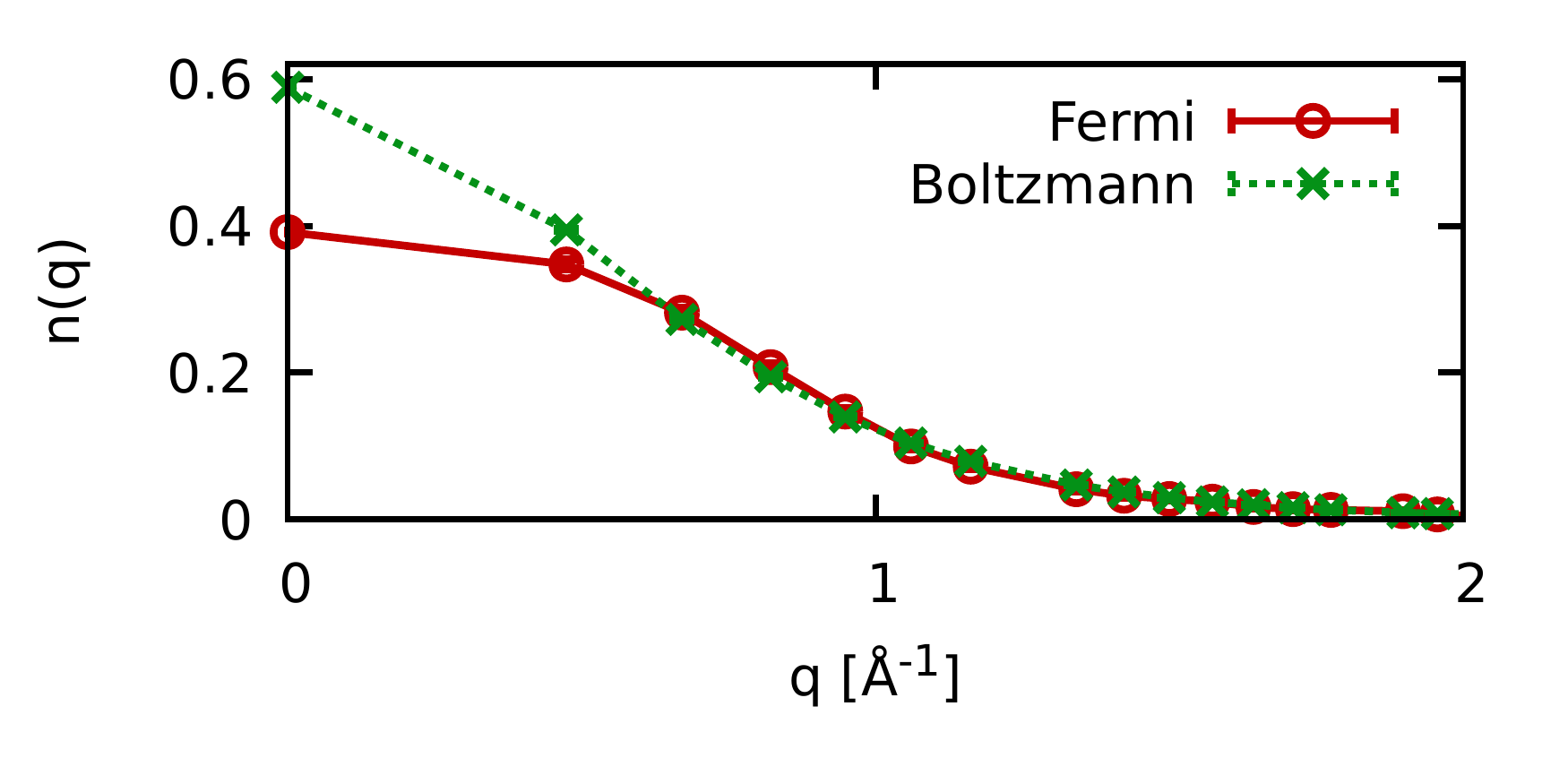}\\\vspace*{-1.23cm}
\hspace*{0.195cm}\includegraphics[width=0.416\textwidth]{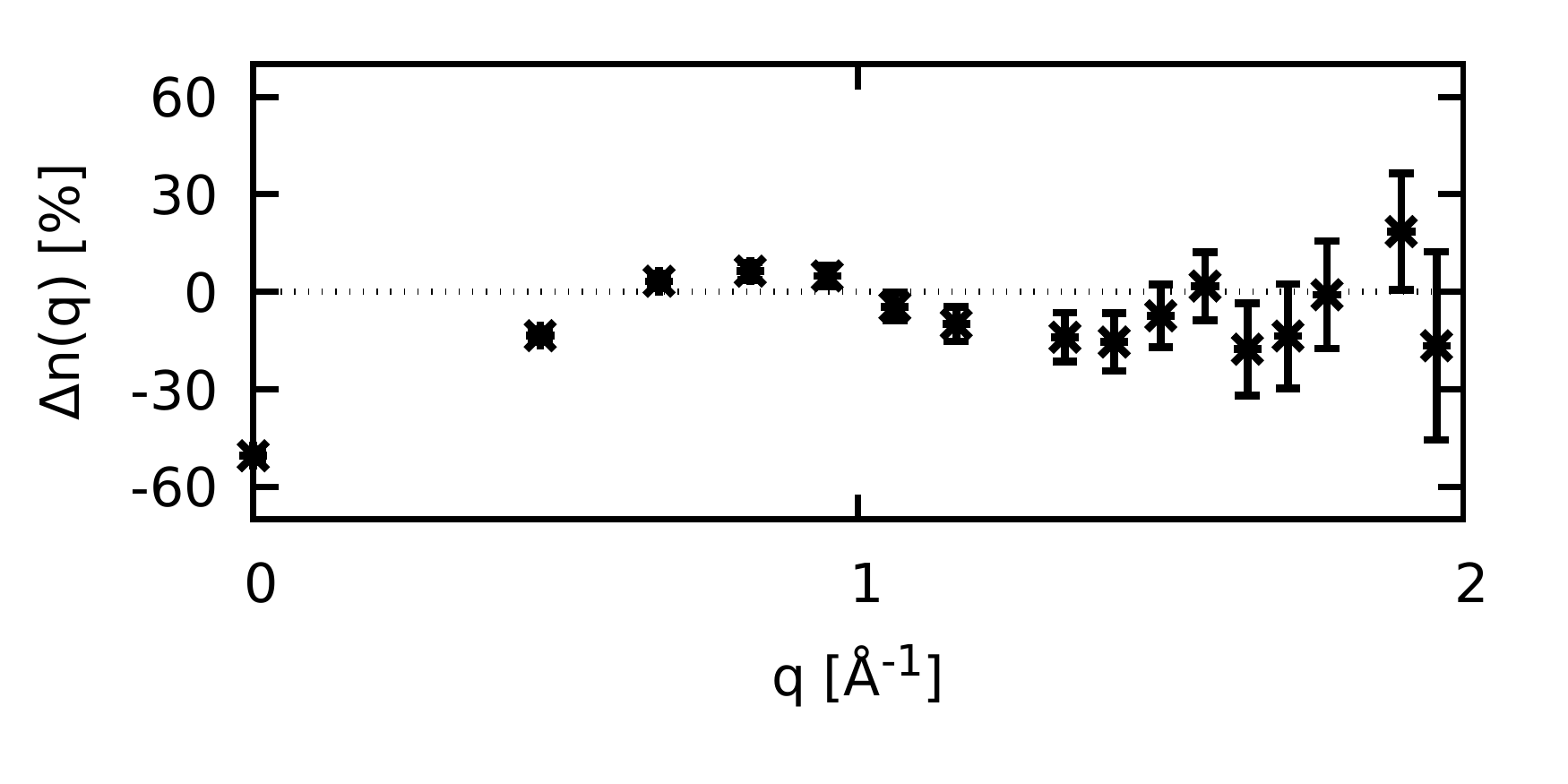}\\\vspace*{-0.5cm}
\caption{\label{fig:Statistics}
Top: momentum distribution $n(q)$ obtained with Fermi and Boltzmann statistics for $T=2$K. Bottom: relative deviation between the two curves. 
}
\end{figure}

Let us next briefly touch upon the impact of quantum statistics on the structural properties of $^3$He, which is shown in Fig.~\ref{fig:Statistics} for the case of $n(\mathbf{q})$ at $T=2$K. Specifically, we compare exact fermionic PIMC results (red) to a corresponding simulation of distinguishable particles at the same conditions, i.e., so-called \emph{boltzmannons}. Evidently, there appear pronounced differences between the two data sets exceeding $50\%$ for the zero-momentum state; see also the bottom panel showing the relative deviation. In stark contrast, the impact of quantum statistics on $S(\mathbf{q})$ and $\chi(\mathbf{q})$ does not exceed $1\%$ at these conditions, which is qualitatively consistent to previous findings for $^4$He~\cite{cep}.

The final property of $^3$He that we investigate in this work is the dynamic structure factor $S(\mathbf{q},\omega)$, which we obtain by numerically inverting the equation~\cite{JARRELL1996133}
\begin{eqnarray}\label{eq:DSF}
F(\mathbf{q},\tau) = \int_{-\infty}^\infty \textnormal{d}\omega\ S(\mathbf{q},\omega) e^{-\tau\omega}\ .
\end{eqnarray}
Specifically, we employ a genetic algorithm similar to the scheme presented in Ref.~\cite{Vitali_PRB_2010}, which simultaneously minimizes the $\chi^2$-measure of Eq.~(\ref{eq:DSF}) and the first and inverse frequency moments; see the Supplemental Material~\cite{supplement} for more details.
To our knowledge, there do not exist experimental measurements of $S(\mathbf{q},\omega)$ in the temperature range $T\in[2,5]$K that is accessible to the present direct PIMC simulations. On the other hand, our investigation of $S(\mathbf{q})$ and $\chi(\mathbf{q})$, both of which are closely related to $S(\mathbf{q},\omega)$, has revealed no significant impact of quantum statistics. Therefore, we have carried out PIMC simulations of $N=100$ $^3$He atoms using Boltzmann statistics at $T=1.2$K, since at this temperature neutron scattering data have been presented by Sk\"old \textit{et al.}~\cite{Skoeld_1980}.


\begin{figure}\centering
\includegraphics[width=0.485\textwidth]{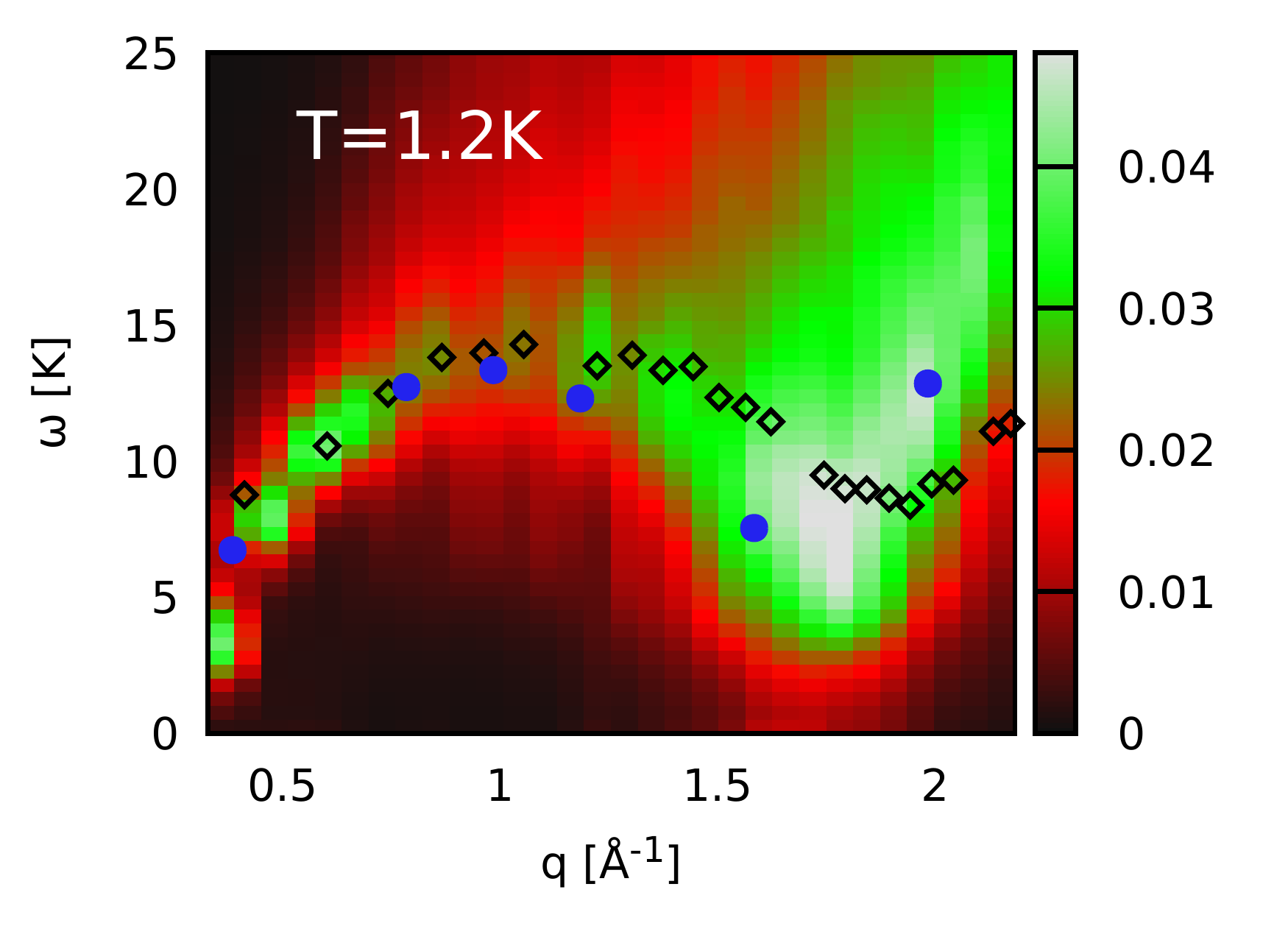}\\ \vspace*{-0.29cm}
\caption{\label{fig:Spec}
Heatmap of reconstructed dispersion $S(\mathbf{q},\omega)$ of $^3$He at $T=1.2$K. Good agreement is found with the experimental data from Ref.~\cite{Skoeld_1980} (blue circles). Because of the higher mass, the roton minimum is shifted to higher $q$ for $^4$He as the theoretical results from Ref.~\cite{Ferre_Boronat_PRB_2016} show (black diamonds).
}
\end{figure}

The results of the numerical inversion of Eq.~(\ref{eq:DSF}) are shown as the heatmap in Fig.~\ref{fig:Spec}. In addition, the blue circles are the experimental peak positions of $S(\mathbf{q},\omega)$ from Ref.~\cite{Skoeld_1980}, and the black diamonds show the same information from a theoretical investigation of $^4$He at the same $T$~\cite{Ferre_Boronat_PRB_2016}. First, we note that all depicted data sets exhibit the phonon-maxon-roton dispersion relation that is well known from $^4$He. Therefore, our results fully corroborate previous findings for $^3$He in two dimensions~\cite{Panholzer_Nature_2012,Nava_PRB_2013}. In addition, we note that the experimental data are in excellent agreement to our results, whereas the $^4$He results substantially deviate in particular for $q\gtrsim1.5$\AA$^{-1}$. Given that we have used Boltzmann statistics in these simulations, this is a strong indication that the spectrum of collective excitations is predominantly shaped by the interaction. The observed differences between the two helium isotopes are, therefore, mainly a mass effect as the heavier $^4$He is more strongly coupled than $^3$He.

\begin{figure}\centering
\includegraphics[width=0.485\textwidth]{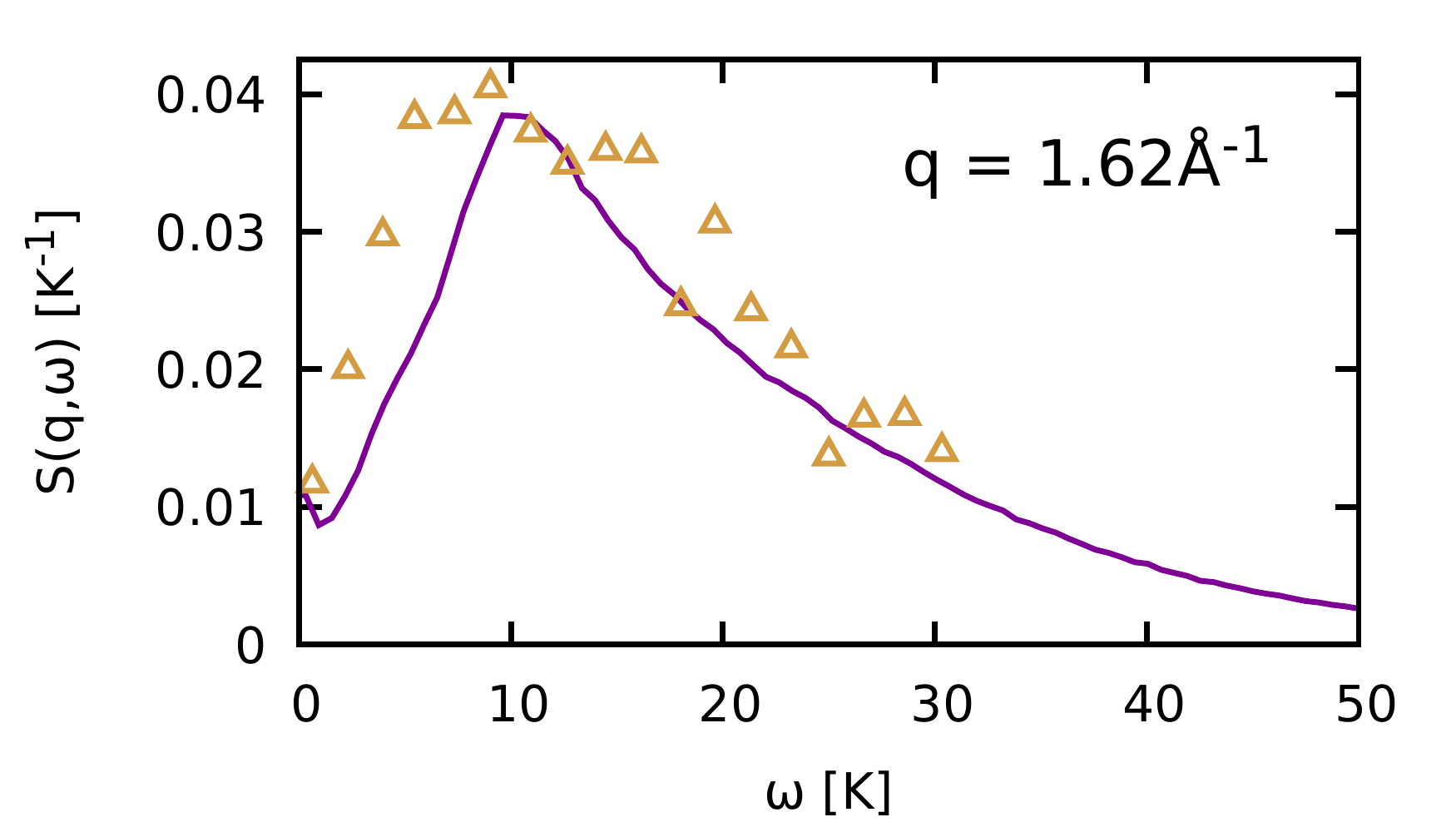}\\\vspace*{-1.03cm}
\includegraphics[width=0.485\textwidth]{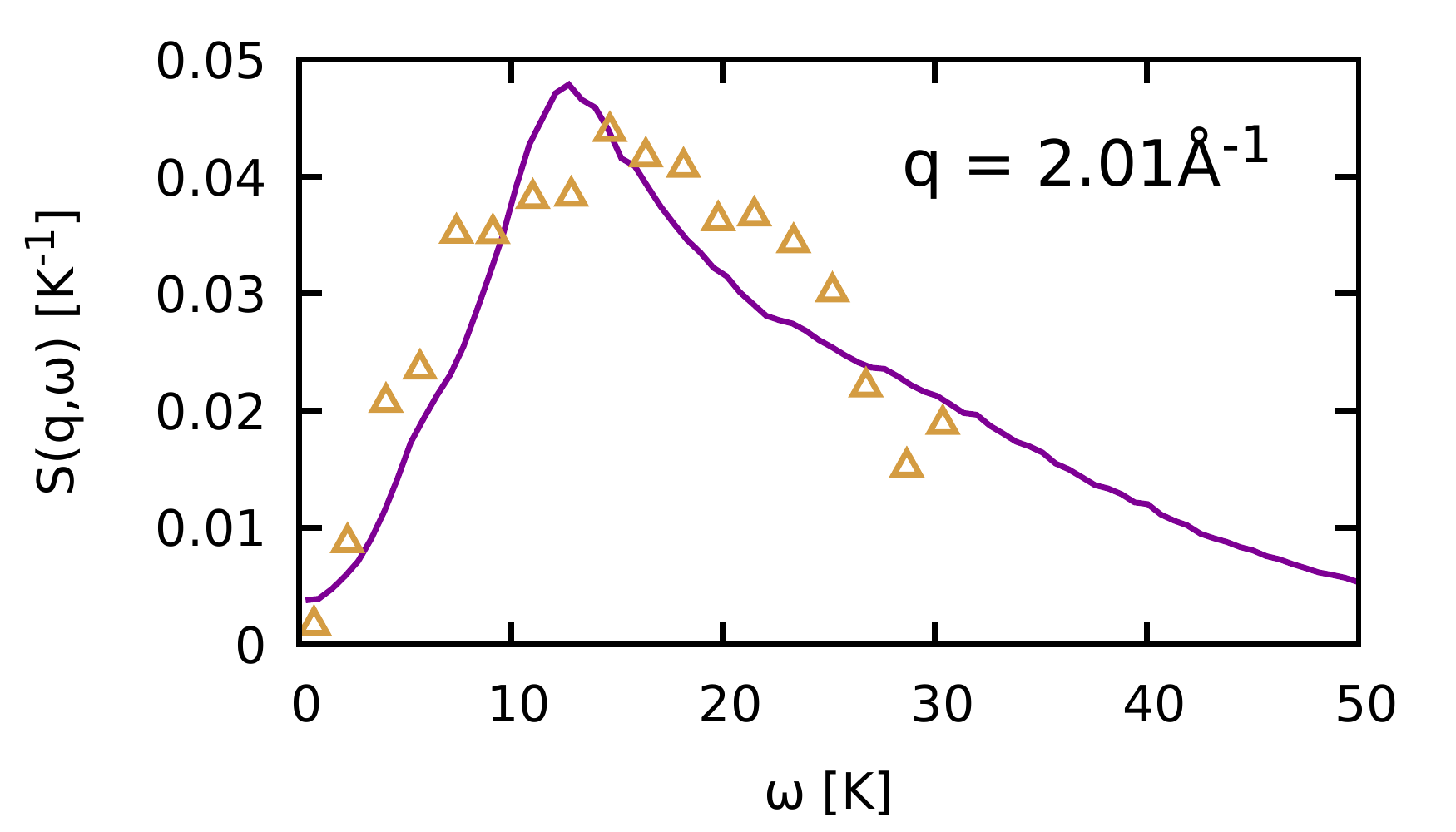}
\caption{\label{fig:Spectrum_T}
Dynamic structure factor $S(\mathbf{q},\omega)$ at $T=1.2$K. Solid purple: reconstructed PIMC data; yellow triangles: experimental data by Sk\"old \textit{et al.}~\cite{Skoeld_1980}.
}
\end{figure}

Let us conclude this study by investigating the full $\omega$-dependence of $S(\mathbf{q},\omega)$, which is shown in Fig.~\ref{fig:Spectrum_T} for $q=1.62$\AA$^{-1}$ [top] and $q=2.01$\AA$^{-1}$ [bottom]. Such comparisons are of fundamental importance to assess the quality of a theoretical method, as the peak position contains comparably limited information~\cite{Ramakrishna_PRB_2021}.
Evidently, the two independent data sets are in excellent agreement over the entire frequency range. This, in turn, means that the PIMC simulation and subsequent reconstruction is not only capable to reproduce the correct dispersion $\omega_\textnormal{max}(q)$, but also gives access to the actual shape of the respective dynamic structure factor $S(\mathbf{q},\omega)$. From a physical perspective, we note that both considered wave numbers are located around the roton minimum [cf.~Fig.~\ref{fig:Spec}], but do not exhibit the sharp quasi-particle excitation peak that is characteristic for the superfluid phase of $^4$He~\cite{Ferre_Boronat_PRB_2016}.

\textbf{Discussion.} In summary, we have presented extensive new \textit{ab initio} PIMC results for normal liquid $^3$He. Specifically, we have carried out direct PIMC calculations for $T\in[2,5]$K, which are computationally challenging due to the fermion sign problem, but are exact within the given Monte Carlo error bars. This has allowed us to obtain the first rigorous theoretical results for different static properties of bulk $^3$He such as the static structure factor $S(\mathbf{q})$, the momentum distribution function $n(\mathbf{q})$, and the static density response function $\chi(\mathbf{q})$. From a physical perspective, we have found that the correlation-induced peak in $\chi(\mathbf{q})$ strongly depends on $T$, whereas $S(\mathbf{q})$ remains almost unaffected. Moreover, the exact nature of our simulations without any nodal restrictions has allowed us to unambiguously quantify the impact of Fermi statistics on these properties, which is quite pronounced for $n(\mathbf{q})$ for small momenta, but practically negligible for $S(\mathbf{q})$ and $\chi(\mathbf{q})$ at these conditions.

Being motivated by this apparent absence of quantum statistics effects on the latter quantities, we have carried out PIMC simulations of $^3$He using Boltzmann statistics at $T=1.2$K, which would otherwise not be feasible due to the sign problem. This has allowed us to compare the dynamic structure factor $S(\mathbf{q},\omega)$ to experimental measurements, and the two data sets are in excellent agreement both regarding the peak position $\omega_\textnormal{max}(q)$ and the actual shape of the respective spectra. In particular, we have found the familiar phonon-maxon-roton dispersion relation that is well known from $^4$He. This substantiates the previous findings for $^3$He in two dimensions~\cite{Panholzer_Nature_2012,Nava_PRB_2013}, where it has been reported that the shape of the dispersion is predominantly shaped by the interaction and not by the type of quantum statistics.

Overall, our new results considerably extend the current understanding of one of the most important and widely studied quantum systems in the literature, which is important in its own right. Future extensions of our work might include the adaption of our set-up to quantum dipole systems and other types of fermionic ultracold atoms. Furthermore, it is, in principle, possible to study $^3$He based on direct PIMC simulations in the grand-canonical ensemble~\cite{Dornheim_JPA_2021}, which would give access to additional interesting physical properties such as the compressibility and the single-particle spectrum $A(\mathbf{q},\omega)$~\cite{Filinov_PRA_2012}.

\begin{acknowledgements}

 We gratefully acknowledge helpful feedback by Paul Hamann on the topic of analytic continuation.\\
This work was partly funded by the Center for Advanced Systems Understanding (CASUS) which is financed by Germany's Federal Ministry of Education and Research (BMBF) and by the Saxon Ministry for Science, Culture and Tourism (SMWK) with tax funds on the basis of the budget approved by the Saxon State Parliament. B.M.~received support from the U.S. Department of Energy (DE-SC0016248).
The PIMC calculations were carried out at the Norddeutscher Verbund f\"ur Hoch- und H\"ochstleistungsrechnen (HLRN) under grant shp00026, on a Bull Cluster at the Center for Information Services and High Performance Computing (ZIH) at Technische Universit\"at Dresden, and
on the cluster \emph{hemera} at Helmholtz-Zentrum Dresden-Rossendorf (HZDR).

\end{acknowledgements}

\bibliography{bibliography}
\end{document}